\documentclass[twocolumn,showpacs,amsmath,amssymb,pra,graphics,graphicx]{revtex4-1}
       \usepackage{graphicx}
        \usepackage{graphics}
\usepackage{bm}
\usepackage{dcolumn}
\begin{document}

\title{Interaction between moving Abrikosov vortices in type-II superconductors}

\author{V. G. Kogan}
\email{kogan@ameslab.gov}
 \affiliation{Ames Laboratory--DOE, Ames, IA 50011, USA}
 \author{R. Prozorov}
\email{prozorov@ameslab.gov}
 \affiliation{Ames Laboratory--DOE, Ames, IA 50011, USA}   
 \affiliation{Department of Physics and Astronomy, Iowa State University, Ames, IA 50011, USA}

  \date{7 April 2020}
       
\begin{abstract}
The  self-energy of a moving vortex is shown do decrease with increasing velocity. The interaction energy of two parallel slowly moving vortices differs from the static case by a small term $\propto v^2$; the ``slow" motion is defined as having the velocity $v<v_c =c^2/4\pi\sigma\lambda$, where $\sigma (T)$ is conductivity of the normal excitations and $\lambda (T)$ is London penetration depth. For higher velocities,  $v>v_c(T)$, the interaction energy of two vortices situated along the velocity direction is enhanced and along the perpendicular direction is suppressed compared to the static case. 
 \end{abstract}

%\noindent \pacs{74.20.-z,74.20.De,74.50.+r}

\maketitle

\section{Introduction}

Recent experiments have tracked Abrikosov vortices moving with  extremely high velocities well exceeding the speed of sound \cite{Eli,Denis}. The time-dependent Ginzburg-Landau equations (GL) was the major tool used to model vortex motion. Although this approach, strictly speaking, is applicable only for gapless systems near the critical temperature \cite{Kopnin-Gor'kov}, it  reproduces qualitatively major features of the fast vortex motion.  

A much simpler  linear  London approach had been successfully employed through the years to describe static or nearly static vortex systems. The London equations express the basic Meissner effect and can be used at any temperature for problem where  vortex cores are irrelevant. The magnetic structure of moving vortices was commonly considered the same as that of a static vortex displaced as a whole.

It has been shown recently, however,  that this is not the case within the Time Dependent London theory (TDL) which takes into account normal currents, a necessary consequence of moving magnetic structure of a vortex \cite{TDL}. In this paper we show  that the line energy of a moving vortex decreases with increasing velocity. Moreover, the interaction of two vortices moving with the same velocity becomes anisotropic so that the interaction is enhanced when the vector $\bm R$ connecting  vortices is parallel the velocity $\bm v$ and suppressed if  $\bm R \perp \bm v$.

\subsection{Outline of Time Dependent London approach}

        In   time dependent situations,   the current   consists, in
general, of  normal and superconducting parts:
\begin{equation}
{\bm J}= \sigma {\bm E} -\frac{2e^2 |\Psi|^2}{mc}\, \left( {\bm
A}+\frac{\phi_0}{2\pi}{\bm
\nabla}\chi\right)  \,,\label{current}
\end{equation}
where ${\bm E}$ is the electric field and  $\Psi$ is the order parameter.  

The conductivity  $\sigma$ approaches the normal state value  $\sigma_n$
when the temperature $T$ approaches $T_c$ in fully gapped s-wave
superconductors; it vanishes  fast with decreasing
temperature along with the density of normal excitations. This is, however, not the case for strong pair-breaking
  when   superconductivity becomes gapless while the density of
states approaches the normal state value at all temperatures. 

Within the London approach $|\Psi|$ is a constant $ \Psi_0 $ and Eq.\,(\ref{current})
reads:
\begin{equation}
\frac{4\pi}{c}{\bm J}= \frac{4\pi\sigma}{c} {\bm E} -\frac{1}{\lambda^2}\,
\left( {\bm A}+\frac{\phi_0}{2\pi}{\bm
\nabla}\chi\right)  \,,\label{current1}
\end{equation}
where $\lambda^2=mc^2/8\pi e^2|\Psi_0|^2 $ is the London penetration depth.
Acting on this by curl one obtains:
\begin{equation}
-\nabla^2{\bm H}+\frac{1}{\lambda^2}\,{\bm
H}+\frac{4\pi\sigma}{c^2}\,\frac{\partial {\bm H}}{\partial
t}=\frac{\phi_0}{\lambda^2}{\bm z}\sum_{\nu}\delta({\bm r}-{\bm r_\nu})\,,\label{TDL}
\end{equation}
where ${\bm r_\nu}(t) $ is the position of the $\nu$-th vortex, $\bm z$ is the direction of vortices that coincides with that of $\bm H$ for isotropic infinite type-II superconductors.
Equation (\ref{TDL}) can be considered as a general form of the time
dependent London equation. This form differs from that provided by F.\,London where contribution of normal
quasiparticles to the current was not included \cite{London}.

As with the static London approach, the time dependent version (\ref{TDL}) has the shortcoming of being valid only outside vortex cores.  As such it may produce useful results for materials with large GL parameter $\kappa$ in fields away of the upper critical field $H_{c2}$. On the other hand, Eq.\,(\ref{TDL}) is a useful, albeit approximate, tool for low temperatures where GL theory does not work and the microscopic theory is forbiddingly complex. 

%%%%%%
\subsection{Moving vortex  }
%%%%%%%

  For a straight vortex along ${\bm z}$ moving with
a constant velocity ${\bm v}$ in the $xy$ plane  Eq.\,(\ref{TDL}) reads:
\begin{equation}
-\lambda^2\nabla^2 H +
H +\tau\,\frac{\partial  H }{\partial
t}= \phi_0 \delta({\bm r}-{\bm v}t)\,,\label{TDL1}
\end{equation}
where $H$ is the $z$ component of the magnetic field and
 \begin{equation}
 \tau= 4\pi\sigma\lambda^2/c^2   
 \label{tau}
 \end{equation}
is the ``current relaxation time", the term used in literature on  time-dependent GL  models.
   Clearly, the field distribution described by Eq.\,(\ref{TDL1}) differs from the   solution which would have existed in the absence of relaxation term  for $\tau=0$:
\begin{equation}
H_0({\bm r},t)=\frac{\phi_0}{2\pi\lambda^2}\,K_0\left(\frac{|{\bm r}-{\bm
v}t|}{\lambda}\right)\,.\label{h0(r,t)}
\end{equation}

 Equation\,(\ref{TDL1}) can
be solved by first finding the time dependence of the Fourier transform
$H_{\bm k}$, as it is done for the diffusion equation \cite{LL}: 
\begin{equation}
\tau\,\partial_tH_{\bm k} + (1+\lambda^2k^2)H_{\bm k}= \phi_0 \,e^{-i{\bm k}
{\bm v} t} 
\end{equation}
which yields
\begin{equation}
       H_{\bm k} =  \frac{\phi_0 \,e^{-i{\bm k} {\bm v} t}}{
1+\lambda^2k^2-i{\bm k}{\bm v}\tau}   \,.
\label{Hk}
\end{equation}
%For a stationary case, the arbitrary $C$ should be set zero.
To find the field distribution in   real space for the stationary case of a constant velocity  one may consider $t=0$. This was done in Ref.\,\onlinecite{TDL} where it was shown that the moving vortex looses the cylindrical symmetry of vortex at rest, in particular, this distribution is no longer symmetric relative to $x\to -x$ with $x$ being the velocity direction.
  
Physically, the distortion of the field distribution is due to contribution
of the out-of-core normal excitations to vortex currents.   At small velocities, the distortion can be disregarded.
  At low temperatures, the quasiparticles are nearly absent (for
the s-wave symmetry) and $\sigma \approx 0$, whereas $\lambda$  is finite, therefore the vortex field distortion is weak. 
Hence, the distortion may have an effect at high $T $s where
the conductivity is close to that of the normal phase. Gapless
superconductors are an exception to this rule, since the normal
excitations density of states is close to the normal even at low $T$s. 

%%%%%%% 
 \section{Self-energy of moving vortex}
%%%%%
 
 Given the field distribution of a moving  vortex, one readily evaluates the London  line energy of a   vortex \cite{deGennes,force}:
\begin{eqnarray}
   F_1 &=&  \int d^2{\bm r}
 \left[ H^2 +\lambda^2( {\rm curl} \bm H)^2\right]/8\pi \nonumber\\
& =& 
 \int \frac{d^2{\bm k}}{32\pi^3} 
 \left[ |H_{\bm k}|^2 +\lambda^2| {\bm k}\times  {\bm H}_{\bm k}|^2 \right] \,, 
 \label{eA1}
\end{eqnarray}
 where the Fourier transform $H_{\bm k}$ is given in Eq.\,(\ref{Hk}) for $t=0$. Further, we have $| {\bm k}\times  H_{\bm k}|^2 = k^2 |H_{\bm k}|^2$, so that
 \begin{eqnarray}
&& \frac{32\pi^3\lambda^2}{\phi_0^2}  F_1 =  \int \frac{d^2{\bm q}\,(1+q^2)}{|1+q^2-i {\bm q}{\bm u}|^2}
   \nonumber\\
&&=  \int \frac{d^2{\bm q}\,(1+q^2)}{(1+q^2)^2+q_x^2 u^2} \,, 
 \label{eA2}
\end{eqnarray}
Here, the dimensionless ${\bm q}=\lambda {\bm k}$ is introduced and the normalized velocity $\bm u=\bm v/v_c$, $v_c=c^2/4\pi\sigma\lambda$ (this $v_c$ is by a factor of 2 smaller than $v_c$ used in Ref.\,\onlinecite{TDL}). After integration over the angle $\varphi$ ($q_x=q\cos\varphi$)
 one obtains the last integral in the form
  \begin{eqnarray}
&&   \int_0^\kappa \frac{2\pi \,q(1+q^2)\,d  q}{\sqrt{(1+q^2)^2+q ^2 u^2} } \nonumber\\
&&   =\pi\ln \frac{2\left(1+\kappa^2+\sqrt{(1+\kappa^2)^2+\kappa ^2 u^2}\right) +u^2 }{4+u^2}\qquad,
\end{eqnarray}
 where the logarithmically divergent integral is truncated at $q=\lambda/\xi=\kappa$. The reduced line energy $f=F_1/(\phi_0^2/32\pi^2\lambda^2)$ for $\kappa=10$ is shown in Fig.\,\ref{fig1}:
 
  \begin{figure}[h]
%\begin{center}
\includegraphics[width=7.5cm] {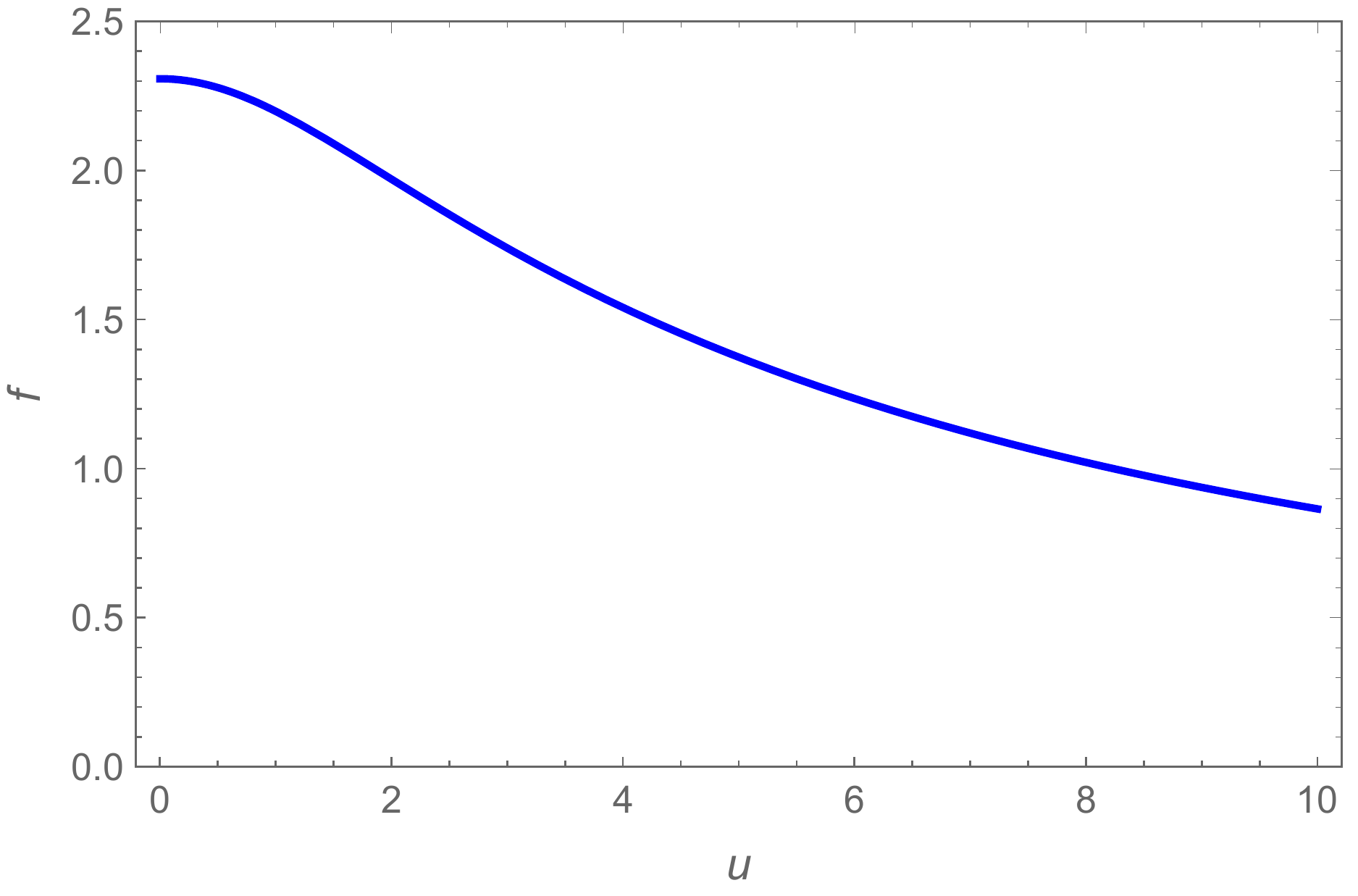}
\caption{(Color online) The line energy $f$ normalized on $\phi_0^2/32\pi^2\lambda^2$ as a function of reduced velocity $u=v/v_c$ for $\kappa=10$. 
}
\label{fig1}
%\end{center}
\end{figure}

It is worth noting that large values of the reduced velocity $u=v/v_c$ are not necessarily imply a large actual velocity because $v_c$ depends on temperature, in particular, $v_c\to 0$ when $T\to T_c$. 
 For a ``fast" motion such that $\kappa^2 \gg u^2\gg 1$, this gives
  \begin{eqnarray}
F_1\approx  \frac{\phi_0^2}{32\pi^2\lambda^2}    \ln \frac{2 \kappa^2  }{ u^2}\,,
\end{eqnarray}
i.e., the line energy is slowly decreases with increasing velocity.

%%%%%%%
 \section{Intervortex interaction}
 %%%%%%
 
 For two parallel vortices  moving with the same velocity, one at the origin at $t=0$  and the other at $\bm R=(x,y)$, the field at $\bm R$ is given by the Fourier transform:
\begin{equation}
       H_{\bm q} =  \frac{\phi_0 (1+e^{-i{\bm q} {\bm R} })}{1+q^2-i q_x u}\,.
\label{Hk2}
\end{equation}
 %where $\lambda$ is used as the unit of length. 
 Using Eqs.\,(\ref{eA1}), one obtains   the total energy $F$ of two vortices and the interaction energy $F_{int}=F-2 F_1$ where $F_1$ is the line energy of a single vortex given in Eq.\,(\ref{eA2}):
 \begin{eqnarray}
  \frac{16\pi^3\lambda^2}{\phi_0^2}  F_{int}({\bm R}) =  \int \frac{d^2{\bm q}\,(1+q^2)\cos {\bm q} {\bm R}}{(1+q^2)^2+q_x^2 u^2}\,.
  \label{eB2}
\end{eqnarray}
For $u=0$ this yields the static interaction energy \cite{deGennes}:
 \begin{eqnarray}
  F_{int} = \frac{\phi_0^2} {16\pi^3\lambda^2} \int \frac{d^2{\bm q}\, \cos {\bm q} {\bm R}}{1+q^2 }=  \frac{\phi_0^2} {8\pi^2\lambda^2}\,K_0\left(\frac{R}{\lambda}\right).\qquad
  \label{eB3}
\end{eqnarray}
Commonly, this energy is written as $ F_{int}=\phi_0 H_{12}(R)/4\pi$, where $H_{12}(R)$ is the field generated by the first vortex  at the location of the second.
\begin{figure}[thb]
\includegraphics[width=7.5cm] {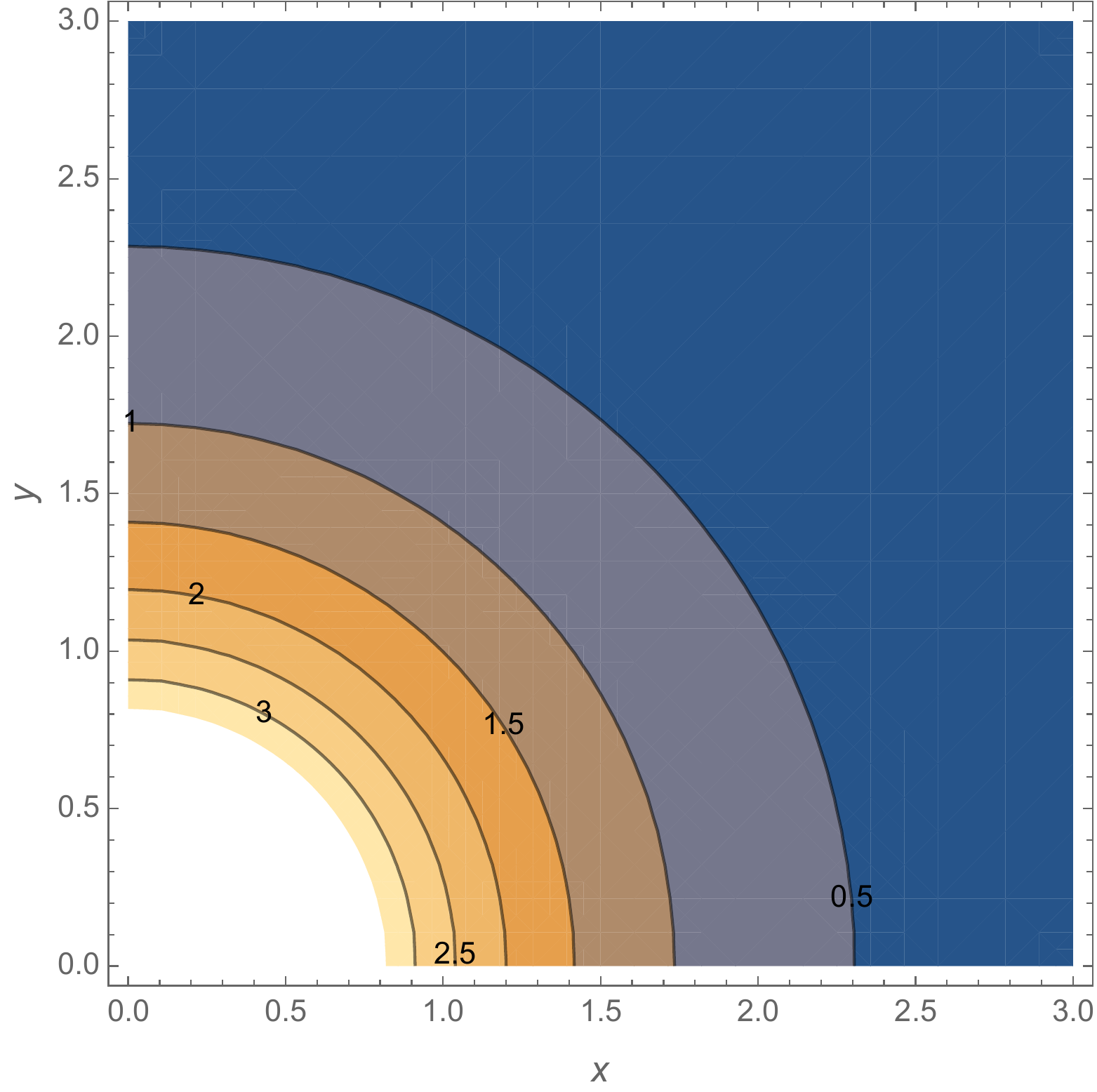}
\includegraphics[width=7.5cm] {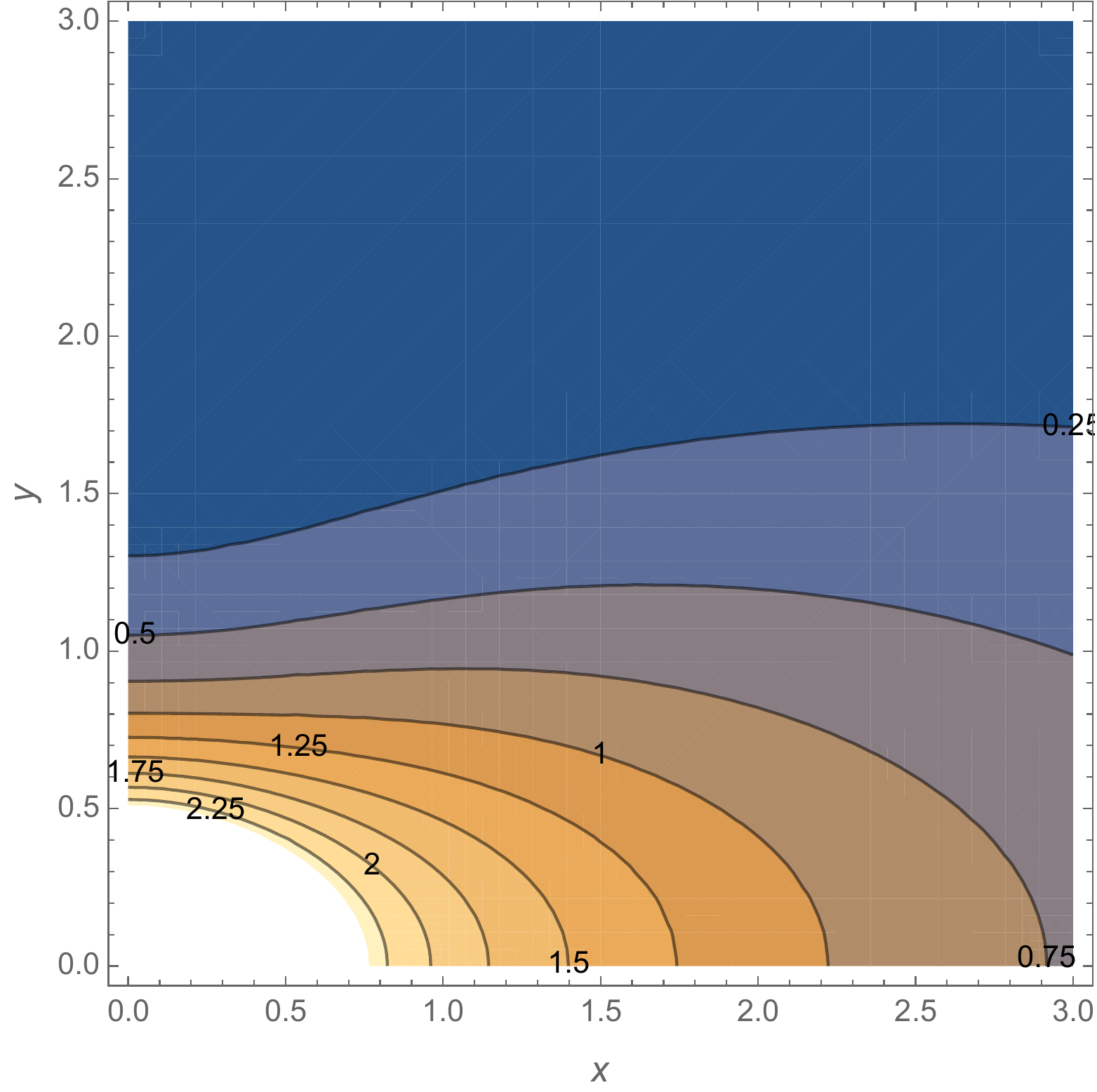}
\caption{(Color online) Contours of constant interaction energy for a vortex at the origin $(0,0)$ and another one at $(x,y)$ for  a small velocity $u=v/v_c=0.2$ (the upper panel) and for fast vortices with $u=v/v_c=4$ (the lower panel).  $x,y$ are measured in units of $\lambda$. In this calculation $\kappa=8$ and $q_m=12$.
}
\label{f2}
\end{figure}

 The double integral (\ref{eB2})  can be evaluated  numerically. 
The integral over $ q$, however, diverges logarithmically. 
One can isolate effects of motion by subtracting and adding the result (\ref{eB3}) for the interaction of vortices at rest:
 \begin{eqnarray}
   f_{int}  = 2 \pi K_0\left( R \right)- \int \frac{d^2{\bm q}\,u^2 q_x^2\cos ({\bm q}  {\bm R})}{(1+q^2)[(1+q^2)^2+q_x^2 u^2]}.\qquad  
  \label{e21}
\end{eqnarray}
where the reduced interaction $ f_{int}=(16\pi^3\lambda^2/\phi_0^2)F_{int} $. Another benefit of this step is that the logarithmic divergence is now incorporated in the exact first term, while the integral here is convergent. 

\begin{figure}[htb]
\includegraphics[width=9cm] {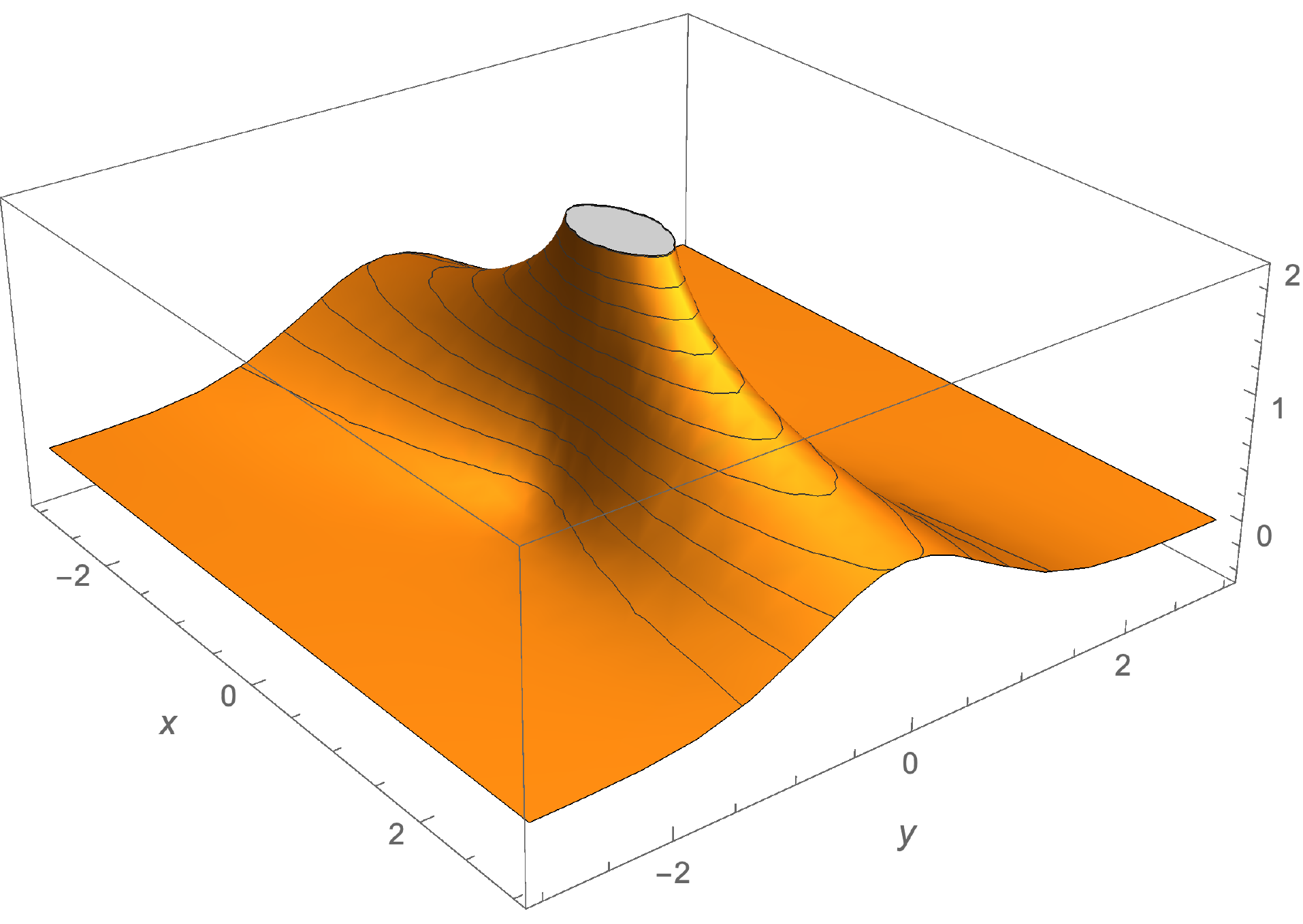}
\caption{(Color online) The   interaction energy $f_{int}(x,y) $ for a vortex at the origin $(0,0)$ and another one at $(x,y)$  both moving along $x$ with  velocity $u=v/v_c=10$. In this calculation $\kappa=10$ and $q_m=20$.
 }
\label{f3}
\end{figure}

To exclude large $|k|>1/\xi$ ($\xi$ is the vortex core size), we introduce a factor $e^{-k^2 \xi^2}=e^{-q^2/\kappa^2}$ in the integrand of Eq.\,(\ref{e21}) and integrate over region $-q_m<q_x<q_m$ and $-q_m<q_y<q_m$ with $q_m$ exceeding $\kappa$ substantially, so that the square shape of the integration domain  
does not matter. 
The result is shown in  Fig.\,\ref{f2}.
% In particular, it is seen that for a fixed intervortex distance $R$, the interaction energy is minimal if the second vortex is situated at $x=0$ and $y=R$. 
The upper panel shows that at low velocities, the interaction energy $F_{int}(x,y)=F_{int}(R,\varphi)$ is nearly azimuth independent. In particular, this means that the interaction force is nearly radial, as is the case of vortices at rest. 
With increasing velocity the situation changes drastically, and the force $-\bm \nabla F_{int}(x,y)$, which is perpendicular to contours  $F_{int}(x,y)$ = const has a complicated distribution.
  To have a better view of the energy $F_{int}(x,y) $ we provide a three-dimensional plot in Fig.\,\ref{f3}.  

Clearly, the interaction energy (\ref{eB2})  remains the same if ${\bm R}\to -{\bm R}$; also, it is symmetric with respect to reflection $y\to -y$. Since the interaction force is $-\nabla F_{int}$,   Fig.\,\ref{f2} shows that the force direction deviates from the direction of $\bm R$, unless $\bm R$ is parallel or perpendicular to the velocity $\bm v=v{\hat{\bm x}}$. 

It is worth noting that the field distribution of the first vortex is asymmetric with respect to $x\to -x$, so that the interaction energy is not proportional to the field of the first vortex at the location of the other.

 %%%%%%%
\section{Electric field and dissipation}
%%%%%%%%

Having the magnetic field (\ref{Hk}) of a moving vortex, one gets for   two vortices of our interest:
\begin{equation}
       H_{\bm q} =  \frac{\phi_0 (1+e^{-i{\bm q} {\bm R} })e^{-iq_xut/\tau}}{1+q^2-i q_x u}\,.
\label{Hk3}
\end{equation}
The moving nonuniform distribution of the vortex
magnetic field causes an electric field $\bm E$  out of the vortex core, which in turn causes the normal currents  $\sigma\bm E$ and the dissipation $\sigma{\bm E}^2$. Usually this dissipation is small relative to Bardeen-Stephen core dissipation \cite{Bardeen-Stephen}, but for fast vortex motion it can become substantial \cite{TDL}. 
 
The field $\bm E$ is expressed in terms of known $\bm H$ with the help of the
Maxwell equations $i({\bm k}\times {\bm E}_{\bm k})_z=- \partial_t H_{z{\bm k}}/c$
and ${\bm k}\cdot{\bm E}_{\bm k}=0$:
\begin{eqnarray}
 E_x &=&  -\frac{\phi_0v}{c} \frac{q_xq_y(1+e^{-i\bm q \bm R  })}{q^2(
1+q^2-iq_x  u )}  \,,\\
E_y&=& \frac{\phi_0v}{c} \frac{q_x^2(1+e^{-i\bm q \bm R  })}{q^2(
1+q^2-iq_x  u )}  \,.
\end{eqnarray}
For the stationary motion, one can consider the dissipation at $t=0$. 

\begin{figure}[htb]
\includegraphics[width=9cm] {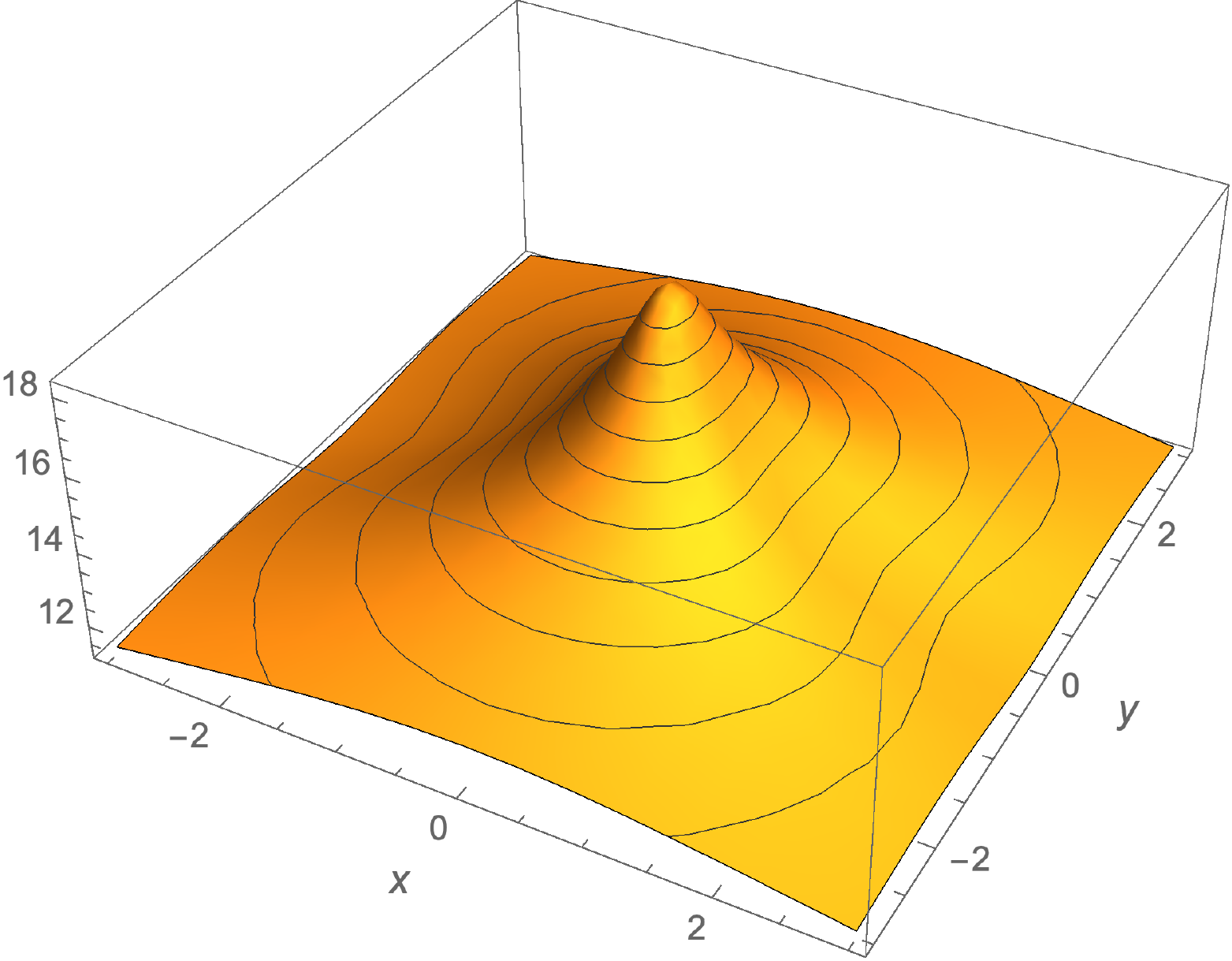}
\caption{(Color online) The reduced  dissipation $w(x,y) $ for a vortex at the origin $(0,0)$ and another one at $(x,y)$  both moving along $x$ with  velocity $u=v/v_c=10$. In this calculation $\kappa=10$ and $q_m=15$.
 }
\label{f4}
\end{figure}

The dissipation power per   unit length is:
\begin{eqnarray}
W=\sigma\int d{\bm r}E^2 =\sigma\int \frac{d^2{\bm
k}}{4\pi^2}\left(|E_{x{\bm k}}|^2+|E_{y{\bm k}}|^2\right)\nonumber\\
=\frac{\phi_0^2\sigma v^2}{\pi^2c^2} \int \frac{d^2{\bm q}\,q_x^2 \cos^2(\bm q \bm R/2)}
{q^2(1+q^2-iq_x u)} \,.
\end{eqnarray}
Treating this integral numerically in the same way as was done for the energy integral in Eq.\,(\ref{e21}), we calculate the reduced quantity $w(x,y)=W (\pi c^2\lambda^2/\phi_0^2\sigma v_c^2)$ shown in Fig.\,\ref{f4}.
 
\begin{figure}[htb]
\includegraphics[width=9cm] {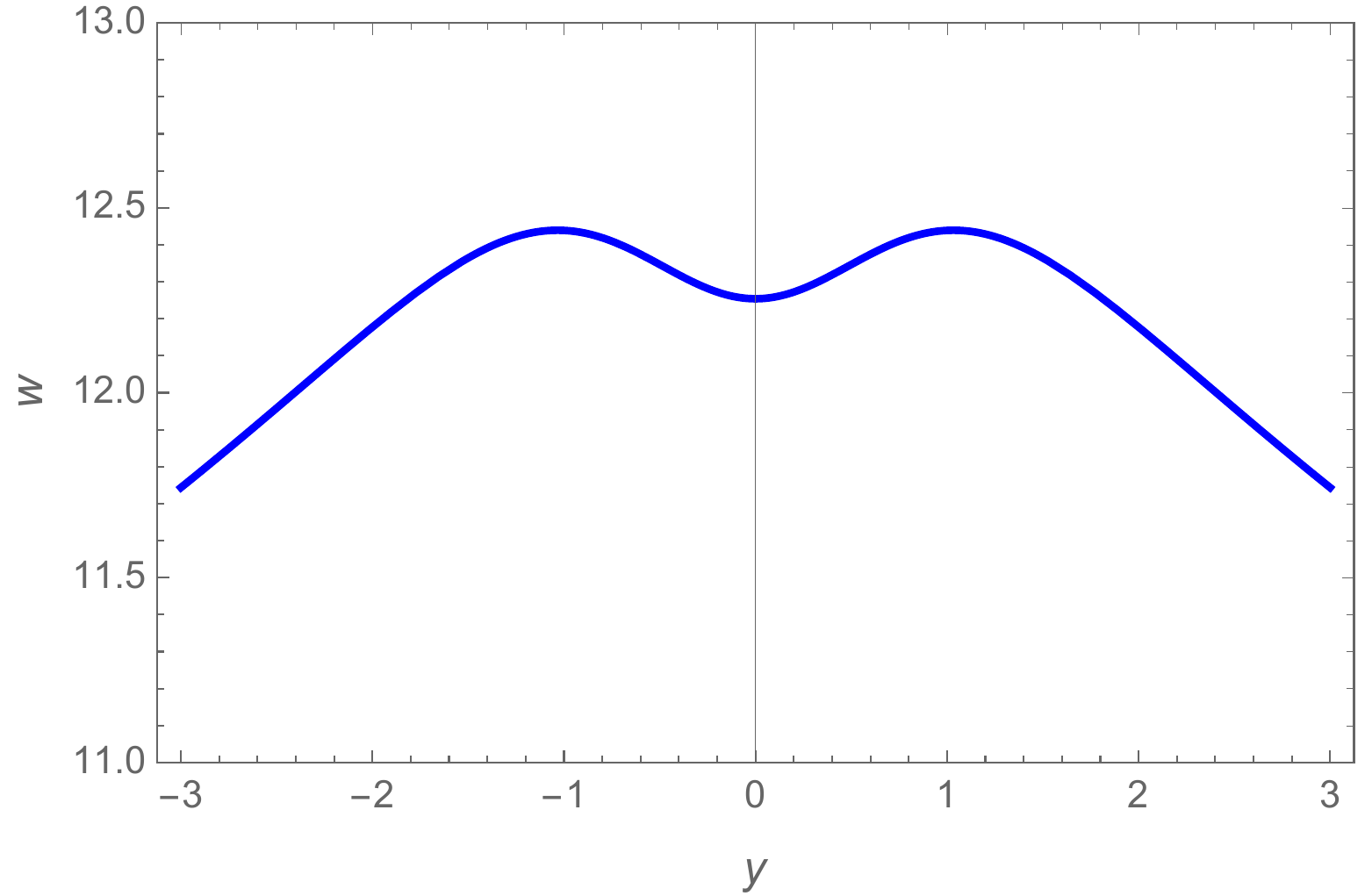}
\caption{(Color online) The reduced  dissipation $w(x,y) $ for a vortex at the origin $(0,0)$ and another one at $(2,y)$ for $u=v/v_c=10$. In this calculation $\kappa=10$ and $q_m=15$.
 }
\label{f5}
\end{figure}

An interesting feature of this result is that the dissipation $w(x,y)$  develops a shallow ditch along the $x$ axis. An example of this ditch   is better seen if we plot a cross section $w(2,y)$ as shown in Fig.\,\ref{f5}. It is seen that for vortices separated by $x\approx 2\lambda$, the ditch width is $\Delta y \approx 2\lambda$, although the dissipation in the minimum is only about 3\% less than at the maxima. 

%%%%%%%
\section{Summary and discussion}
%%%%%%%

The time-dependent London equations, formulated to include normal currents around a moving vortex, show that the vortex field distribution  differs from a static distribution displaced as a whole \cite{TDL}. We argue that 
the self-energy of a moving  vortex is reduced as compared to  the static case and decreases with increasing velocity. 

Moreover, the interaction energy of two vortices, moving with the same velocity, one at the origin at $t=0$ another at $\bm R=(x,y)$,  is symmetric relative to $x\to -x$ ($x$ is along the velocity) notwithstanding the asymmetric field distribution of the first. In other words, the common rule stating that the interaction energy of two vortices is proportional to the field of the first vortex at the location of the other   holds only for the vortices  at rest. 

% ********  About $v_c$ and $u$.*********
 
 As in any London based approach, our results are applicable only out of the vortex cores. The only relevant parameter of the theory, in addition to penetration depth $\lambda$, is the reduced vortex velocity $u=v/v_c$ with  $v_c=c^2/4\pi\sigma\lambda$;   $u$ is usually small away of $T_c$ since the conductivity $\sigma$ of the normal quasiparticles disappears along with the density of normal excitations. In other words, at low temperatures $v_c$ is large, $u$ is small, and effects we discuss here are weak. This is not the case near $T_c$ where 
$v_c\to 0$. Also in the presence of pair breaking, the density of states might be close to that of the normal state at all temperatures  under $T_c$ (gapless case), and the reduced velocity $u$ can be large even for actual velocities  $v$ being relatively moderate.

In experiments  \cite{Eli,Denis}, at velocities exceeding $10^6\,$cm/s, vortices are reported to form chains along the velocity. The moving vortex core has a tail of suppressed order parameter in the $-\bm v$ direction which at large enough velocities may cause the following vortex to trail the first one. 
  A moving vortex generates heat due to normal currents and changing in time order parameter. 
  This complicated process is  discussed  in  \cite{Denis,Eli} in the frame of the time dependent GL theory. 
  
  In this paper we consider a less ambitious and simple model of Abrikosov vortices moving with a constant velocity within time-dependent linear London theory. Whereas distances $\sim\xi$ are unaccessible within this approach, the  interaction of vortices at distances of the order of $\lambda\gg\xi$ are well described by the London - type theory. 
We show that usual models which treat moving vortices as a static, only  displaced in space and time by Galilean transformations, miss nontrivial changes in the vortex field structure and in the intervortex interaction, which, as we show, become relevant for fast motion.

 This work was supported by the U.S. Department of Energy (DOE), Office of Science, Basic Energy Sciences, Materials Science and Engineering Division.  Ames Laboratory  is operated for the U.S. DOE by Iowa State University under contract \# DE-AC02-07CH11358.

\end{document}